\documentclass[12pt,thmsa]{article}
\usepackage{amsfonts}
\usepackage{graphicx}
\usepackage{epsfig}
\usepackage{graphics}

\makeatletter
\newcommand{\row}[1]{\mathord{\buildrel{\lower3pt\hbox{$\scriptscriptstyle\rightarrow$}}\over #1}}

\newcommand{\dyadic}[1]{\mathord{\dyadic@rrow{#1}}}
\newcommand{\dyadic@rrow}[1]{
\begin{picture}(12,12)(-1,0)
\put(-2,12){\makebox(0,0)[t]{$\scriptscriptstyle\downarrow$}}
\put(-2,12){\makebox(0,0)[l]{$\scriptscriptstyle\longrightarrow$}}
\put(5,0){\makebox(0,0)[b]{$#1$}}
\end{picture}
}
\newcommand{\bra}[1]{\bigl\langle #1 \bigr|}
\newcommand{\ket}[1]{\bigl| #1 \bigr\rangle}

\input{tcilatex}
\begin{document}
\begin{center}
 {\Large Commnication via Quantum Neural Network}


{\large{A. Al- Segher$^{(1)}$ and Nasser Metwally$^{(1,2)}$ }}

{\large {\footnotesize $^{(1)}$ Mathematics Department, Faculty of
Science,
Aswan Egypt\\
$^{(2)}$Math. Dept., College of Science, Bahrain University, 32038
Bahrain }}
\end{center}

\begin{abstract}

In this study, the partially entangled  neural networks is used to
transfer information between two neurons, where the original
teleportation protocol is employed this for this  purpose. The
effect of the  network strength on the fidelity of the transported
information is investigated. We show  that as  the strength  of
the network increases, the accuracy of the transformed information
increases. As a practical application, we consider the spread of
swine flu virus between two equivalent tranches of the community.
In this treatment two factors are considered, one for humanity and
the other for influence factor.
 The likelihood of infection between different age group is investigated,
 where we show that the strength of the neural network and the degree of infection plays an important role on transferring infection between different age group.
From theoretical point of view, we show that it is possible to
control  the spread of the virus by controlling  the network
parameter. Also, by using local rotation, one can decrease the
rate of infection between the young.
\end{abstract}



\section{Introduction}

Quantum information, QI is one of the most important events which
has considerable progress theoretically and experimentally. To
manipulate QI, as sending, coding, etc, one needs entangled
quantum channel for these proposes. There are several efforts that
has been done to generate entangled channels. These channels have
been demonstrated in many different types of systems, such as
photons \cite{Aspt}, ions \cite{Blatt}, atoms \cite{jun} and
charged qubit\cite{metwally0}. All these types of quantum channels
have been used to perform different quantum information tasks as
quantum teleportation \cite{Bennt,Olm}, dense coding,
cryptography.

Natural computing is a computational version of the process of
extracting ideas from nature to develop computational systems to
perform computations \cite{Lean}. One of the most promising type
of quantum channels are the neural networks, NN. This field
contains several basic ideas, as  concept of processing elements
(neurons), which represent the correlated processor units, the
transformation performed by these elements and the network
dynamics \cite{kouda,Gral}. Due to its potential in describing the
brain function, the NN has attracted much attentions, where it
represents a model of parallel distributed memory \cite{Amit,Maz}.
In fact NN, can compute any computable functions, so it is
classified as neurocomputer. Also, NN is useful for classification
and pattern recognition \cite{Lean}.

Quantum neural network, QNN represents a type of artificial NN, where it has
all the features of  NN \cite{Alt}. It is one of a typical approaches to
study the human brains, where the information processing in the brain is
mediated by the dynamics of a large interconnected neuronal populations. On
the other hand,  the basic components of neural
cytoskeleton are very likely to possess quantum mechanical properties due to
their size and structure \cite{Pens}. As an example the tubuline protein has
the ability to flip from one conformation to another. These two
conformations act as two basis states of the system. Also the system can lie
in a superposition of these two states, which can give a plausible mechanism
for creating a coherent state in the brain \cite{Beh}.

There are different models that have been considered for QNN. As
an example, in \cite{Mori} the authors have proposed a model  that
is a multilayered neural network and investigated its
characteristic features, such as the effect of quantum
superposition and probabilistic interpretation. A model of
recurrent quantum neural network has been introduced by Behera et
al \cite{Beh}, where they used a linear neural circuit to set up
the potential field in which the quantum brain is dynamically
excited. Also a theoretical quantum brain has been proposed using
a nonlinear schr\"{o}dinger wave equation \cite{Beh1}. A learning
method has been proposed for quantum neural network by Kingo et
al. \cite{kingo}. Shafee \cite{Shafee} has suggested another model
by replacing the classical neurons by quantum bits and quantum
operators in place of the classical action potentials observed in
biological contexts. Pons at. el. \cite{Pons}, have suggested a
model of QNN, where they used the trapped ion chain as quantum
neural network. Based on the universality as single qubit rotation
gate and two-qubit controlled -Not gate, Panchi and Shiyong have
constructed a quantum neuron model \cite{Li}.

Communication at long distances in unknown environments often requires
combatting noise that may affect or destroy data before it has reached to  its
intended destination. It is well known that, entanglement provides miracle
performances in the transmission of quantum information, where it enables
the transmission of quantum state by sending only classical information \cite%
{Bennt}. Recently, Hayashi \cite{Hay} investigated the effect of
the prior entanglement on quantum network coding. This leads us to
our aim; Investigating the effect of non-maximal entangled state (
due to the noise the maximum entangled states turn into partially
entangled states) on the efficient of sending quantum state via
quantum teleportation through a neural quantum network.

In this paper, we consider a partially  entangled quantum neural
network, QNN where each neuron entangled with its neighbor. This
entangled networks between neurons are used to send information
from one neuron to another. This paper is organized as follows: In
Sec.2,  the properties of the QNN  are reviewed. Sending
information between neurons is achieved by quantum teleportation
is described in Sec.3.  The spread of swine flu virous, as an
application of quantum teleportation via neural network, is
discussed in Sec.4. Finally Sec.$5$, is devoted to discuss our
results.

\section{Description of QNN}

A neuron is defined as a two logical state of usual bit "Yes, No"
or" False, true" or "of, on" and so on. Also, one can express the
state of the neuron as a superposition of the False and true
representation as
\begin{equation}  \label{neuron}
\bigl| \psi \bigr\rangle=\alpha\bigl| n \bigr\rangle+\beta\bigl| y %
\bigr\rangle,
\end{equation}
where with probability $|\alpha|^2$ the neuron is in state $\bigl| n %
\bigr\rangle$ i.e "no" has information(unfired), while it is in state $%
\bigl| y \bigr\rangle$ i.e "Yes" carries information (fired) with
probability $|\beta|^2$ and $|\alpha|^2+\beta|^2 = 1$
\cite{Shafee}. In QNN, a neuron receives $N_{i}$ real input
signals from many other neurons by network channel. Also, the
output of a neuron can "fan-out", the number of the input signals
that can be driven by a single output, to several other neurons.
So, we can express on the collective response of all neurons as
\begin{equation}
\bigl| \psi \bigr\rangle=\sum_{i=0}^{N}{\omega_i \bigl| \psi_i \bigr\rangle},
\end{equation}
where $\bigl| \psi_i \bigr\rangle$, represent the states of the
input neurons and $\omega_i$ is the corresponding weight. The
neural network processing can be implemented by an operator
$\mathcal{U}$, acting on the input states and propagates the
information forward to calculate the output state of the neuron as

\begin{equation}
\bigl| \psi(t) \bigr\rangle=\mathcal{U}\Bigl(\sum_{i=0}^{N}{\omega_i \bigl| %
\psi_i \bigr\rangle}\Bigr),
\end{equation}
where, $\mathcal{U}$ is unknown operator that can be implemented
by the gate of the neural network. On the other hand, in quantum
computation the
evolutionary operators must be unitary. If we consider the case where $%
\mathcal{U}=I$, i.e, identity operator, the output of quantum
perceptron is given by
\begin{equation}
\bigl| \psi(t) \bigr\rangle=\sum_{i=0}^{N}{\omega_i \bigl| \psi_i %
\bigr\rangle}.
\end{equation}
This means that quantum preceptron can be defined as a quantum
system consists of n-neuron input register and  in this case, we
have only considered quantum neural network. To achieve quantum
computation, the logical function can be performed by applying a
group of unitary operations on the neuron states. These operations
are called logic gates, which transform the information of the
input neuron states to the output ones. The most basic universal
gates is the single rotation gate, controlled not gate (cNOT),
entangling gate and Hadamard gate. Since we are interested to
employ the QNN into  the quantum teleportation, we shell describe
the cNOT and hadamard gates in our notations. The effect of the
cNOT operation is defined by,

\begin{eqnarray}  \label{CNOT}
cNOT\bigl| nn \bigr\rangle&=&\bigl| nn \bigr\rangle,\quad cNOT\bigl| ny %
\bigr\rangle=\bigl| ny \bigr\rangle,\quad  \nonumber \\
cNOT\bigl| yn \bigr\rangle&=&\bigl| yy \bigr\rangle,\quad cNOT\bigl| yy %
\bigr\rangle=\bigl| yn \bigr\rangle,\quad
\end{eqnarray}

while Hadamard is,
\begin{equation}  \label{Had}
H\bigl| n \bigr\rangle=\frac{1}{\sqrt{2}}(\bigl| n \bigr\rangle+\bigl| y %
\bigr\rangle),\quad H\bigl| y \bigr\rangle=\frac{1}{\sqrt{2}}(\bigl| n %
\bigr\rangle-\bigl| y \bigr\rangle)
\end{equation}

\section{Quantum teleportation}

In this section, the entangled neural network is used  to perform the original
quantum teleportation protocol \cite{Bennt}. To achieve this task we assume
that the neural network is built on the postulates given by Shiafee \cite%
{Shafee}. In this consideration the neuron is represented by
qubit, such that the state of the neuron is described by
equation(\ref{neuron}). Let us assume that the neural network is
partially entangled. This means that the state of each two
neighboring neurons is non maximal entangled states. Assume that
the state of any two connected neurons is defined by the density
operator,
\begin{equation}  \label{nerstate}
\rho=\frac{1+q}{2}\bigl| nn \bigr\rangle\bigl\langle nn \bigr|+\frac{p}{2}(%
\bigl| nn \bigr\rangle\bigl\langle yy \bigr|+\bigl| yy \bigr\rangle%
\bigl\langle nn \bigr|)+\frac{1-q}{2}\bigl| yy \bigr\rangle\bigl\langle yy %
\bigr|,
\end{equation}
where $q^2+p^2=1$ and $0\leq p\leq 1$ and $0\leq q\leq 1$(see
\cite{metwally}).
If, we put $p=1$, one gets the maximum Bell state $\bigl| \phi^+ \bigr\rangle%
\bigl\langle \phi^+ \bigr|$ \cite{Werner}. The following describes the steps
to implement the Bennett Protocol \cite{Bennt}.

\begin{enumerate}
\item \textit{Step one}: A neuron $\mathcal{B}$ has got
information from another neuron $\mathcal{A}$ (which can be order,
computing, decision, etc..). The aim of neuron $\mathcal{B}$ is
transferring  these information to another neuron $\mathcal{C}$,
which is a member in the network. Let us assume
that the unknown information coded in the neuron state $\bigl| \psi_{%
\mathcal{A}} \bigr\rangle=\alpha\bigl| n \bigr\rangle+\beta{\bigl| y %
\bigr\rangle}$ and the neurons $\mathcal{B}$ and $\mathcal{C}$
share an entangled neuron state defined by
$\rho_{\mathcal{BC}}$(\ref{nerstate}),
then the total state of the system is given by $\rho_{\mathcal{S}}=\rho_{%
\mathcal{A}}\otimes\rho_{\mathcal{BC}}$, where $\rho_{\mathcal{A}}=\bigl| %
\psi_{\mathcal{A}} \bigr\rangle\bigl\langle \psi_{\mathcal{A}} \bigr|$.

\item \textit{Step two:} Neuron $\mathcal{B}$ performs  local
operations on his state and the neuron $\rho_{\mathcal{A}}$. These
local operations include the cNOT operation which is a two neurons
gate as defined by (\ref{CNOT}). After this operation one gets
\begin{equation}
\rho^{(1)}_{\mathcal{S}}=(cNOT\otimes I) \rho_{\mathcal{S}} (I\otimes cNOT),
\end{equation}
where $I$, is a unitary operator effects on the neuron
$\mathcal{C}$. The second operation is the Hadamard operation
(\ref{Had}), which transform the state $\rho^{(1)}_{\mathcal{S}}$
to
\begin{equation}
\rho^{(2)}_{\mathcal{S}}=(\mathcal{H}\otimes I\otimes I)\rho^{(1)}_{\mathcal{%
S}}(I\otimes I\otimes \mathcal{H}).
\end{equation}

\item \textit{Step three}: Alice makes  measurements on the basis $\bigl| nn %
\bigr\rangle, \bigl| ny \bigr\rangle,\bigl| yn \bigr\rangle$ and $\bigl| yy %
\bigr\rangle$ randomly on her own neuron and the state which has
got from the neuron $\mathcal{A}$. Alice gets a two classical bit
of information and sends them to Bob.

\item {\it Step four}: As soon as Bob gets the classical data from
Alice, he performs  suitable rotations on his neuron to obtain the
final  information as  shown in Table$(1)$.
\begin{table}[tbp]
\begin{tabular}{|c|c|c|}
\hline
Alice Deuity & Bob state $\rho_b=$ & Bob Deuity \\ \hline
$\ket{nn}\bra{nn}$ & $\frac{|\beta|^2\lambda_1}{2}\ket{nn}\bra{nn}+\frac{\alpha\beta^*\lambda_3}{2}\ket{n}\bra{y}
  +\frac{\beta\alpha^*\lambda_3}{2}\ket{y}\bra{n}+ \frac{|\alpha|^2\lambda_2}{2}\ket{y}\bra{y}$ & $I\rho_b I$ \\ \hline
$\ket{ny}\bra{ny}$ & $\frac{|\beta|^2\lambda_1}{2}\ket{n}\bra{n}-\frac{\alpha\beta^*\lambda_3}{2}\ket{n}\bra{y} - \frac{\beta\alpha^*\lambda_3}{2}\ket{y}\bra{n}+ \frac{|\alpha|^2\lambda_2}{2}\ket{y}\bra{y}$ & $\sigma_x\rho_b\sigma_x$ \\ \hline
$\ket{yn}\bra{yn}$ & $\frac{|\alpha|^2\lambda_1}{2}\ket{n}\bra{y}+ \frac{\alpha\beta^*\lambda_3}{2}\ket{y}\bra{n}+\frac{\beta\alpha^*\lambda_3}{2}\ket{y}\bra{y}+ \frac{|\beta|^2\lambda_2}{2}\ket{y}\bra{y}
\bigr|$ & $\sigma_z\rho_b\sigma_z$ ~ \\ \hline
$\ket{yy}\bra{yy}$ & $\frac{|\alpha|^2\lambda_1
}{2}\ket{n}\bra{n}- \frac{\alpha\beta^*
\lambda_3}{2}\ket{n}\bra{y}
  -\frac{\beta\alpha^*\lambda_3}{2}\ket{y}\bra{n} 
\bigr|+ \frac{|\beta|^2\lambda_2}{2}\ket{y}\bra{y}$ & $\sigma_y\rho_b\sigma_y$ \\ \hline
\end{tabular}
\caption{The possible measurements (Alice side), obtained result
(Bob side) and the local operations which done by Bob to get the
desired coded information, where $\lambda_1=\frac{1+q}{2}, \lambda_2=\frac{1-q}{2}$ and $\lambda_3=\frac{p}{2}$ }
\end{table}
\end{enumerate}

In Fig.1, we plot the fidelity, $\mathcal{f}$ of the teleported state, where we assume the
case where Alice measures  $\bigl| yy \bigr\rangle\bigl\langle yy %
\bigr|$.
\begin{figure}[b!]
\begin{center}
\includegraphics[width=20pc,height=15pc]{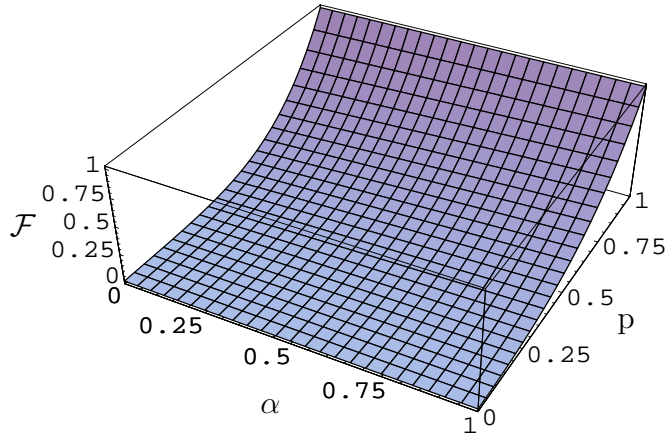}
\put(-155,18){$\alpha$} \put(-20,50){p}
\put(-250,85){$\mathcal{F}$}
\end{center}
\caption{The Fidelity of the teleported state $\mathcal{F}$ a gainst the
noise parameter $p=\protect\sqrt{1-q^2}$ and $\protect\alpha= \protect\sqrt{%
1-\protect\beta^2}$. }
\end{figure}
From this figure, it is clear that when  $p=0$, then  the quantum
network is classically correlated and given by $\rho=\bigl| nn
\bigr\rangle\bigl\langle nn \bigr|$. As one increases the value of
the parameter $p$ on the expanse of $q$, where $p=\sqrt{1-q^2}$,
the quantum network becomes a partially entangled. The
corresponding fidelity, $\mathcal{F}$ increases as $p$ increases.
For $p=1$, the network is maximally entangled  and given by
$\rho=\frac{1}{2}(\bigl|
nn \bigr\rangle\bigl\langle yn \bigr|+\bigl| yn \bigr\rangle\bigl\langle nn %
\bigr|$ and  the fidelity of the teleported state $\mathcal{F}=1$ (maximum
value). On the other hand, there is no effect on the type of the teleported
state on the degree of the fidelity. So, one can teleporte classical
information as well as quantum information with the same efficiency. Then
 by controlling  the devises which generate entangled
network, one can send the information with high fidelity.

\section{\protect\bigskip Application:The spread of swine flu virus}

The prevalence of swine virus is one of the most difficult
challenges facing the world these days. In this contribution, we
use the QNN, which has been described in  Sec.$(2)$ to investigate
the possibility of controlling the spread of this virus. In the
present protocol, we consider two categories of the populations
with the same social and economic circumstances, and each of them
contains  different age groups. Let's assume that each one of
these category is described by a pair of information: the first
part of which represents human characteristics such as young, old,
rich, poor, have a medical culture and no medical culture. The
second part represents information on the  affecting factors  such
as: non-infected, infected, awareness and non awareness, has
culture of health and has no culture of health). In fact, there
are a lots of human and influence factors, so we consider some of
them to illustrate the basic idea of this procotol.

Let us assume that the human properties are codded in the phase, while the
influence factors coded in the amplitude. So, if the neuron is described by $%
\bigl| \psi \bigr\rangle=\alpha\bigl| 1 \bigr\rangle+\beta\bigl| 0 %
\bigr\rangle$, then with probability $|\alpha|^2$, the neuron
represents the pair (young, non-infected), while with probability
$|\beta|^2$, it represents the pair (young, infected). In this
description, we set the positive phase ($+$) for the young and
$(-)$ for the old and the amplitude $1$ for non-infected and $0$
for infected. Table$(2)$, gives a complete description to some of
these factors.

\begin{table}[htp!]
\begin{center}
\begin{tabular}{|c|c|c|c|}
\hline
Human factor & phase & Influence factors & amplitude \\ \hline
young & $+$ & non-infected & $1$ \\
old & $-$ & infect\quad & $0$ \\
rich & $+$ & awareness & $1$ \\
poor & $-$ & non-awareness & $0$ \\ \hline
\end{tabular}%
\end{center}
\caption{This table is an example of coding the information in a
neuron.}
\end{table}

\begin{table}[htp!]
\begin{center}
\begin{tabular}{|c|c|c|c|}
\hline\hline
Age group & Healthy case & Probability & Measurments \\ \hline
old & Infected & $|B_1|^{2}$ &  \\
old & non-infected & $|A_1|^{2}$ & $\bigl| yy
\bigr\rangle\bigl\langle yy \bigr|$
\\
young & Infected & $|B_2|^{2}$ &  \\
young & non-infected & $|A_2|^{2}$ &  \\ \hline
young & Infected & $|B_3|^{2}$ &  \\
young & non-infected & $|A_3|^{2}$ &  \\
old & Infected & $|B_4|^{2}$ & $\bigl| yy \bigr\rangle\bigl\langle
yy \bigr|$
\\
old & non-infected & $|A_4|^{2}$ &  \\ \hline
\end{tabular}%
\end{center}
\caption{This table shows the age group and the possibility of
non-infected or infected.}
\end{table}

\begin{figure}
\begin{center}
\includegraphics[width=18pc,height=12pc]{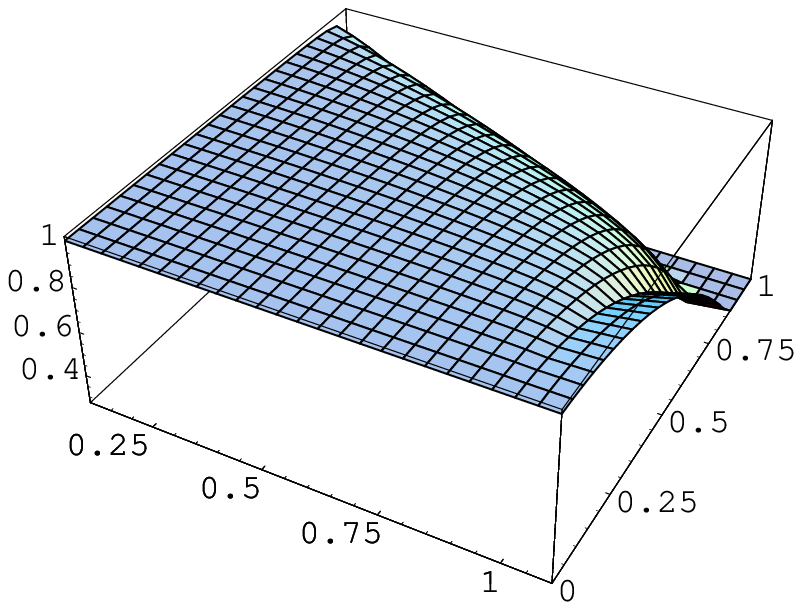}~\quad
\includegraphics[width=18pc,height=12pc]{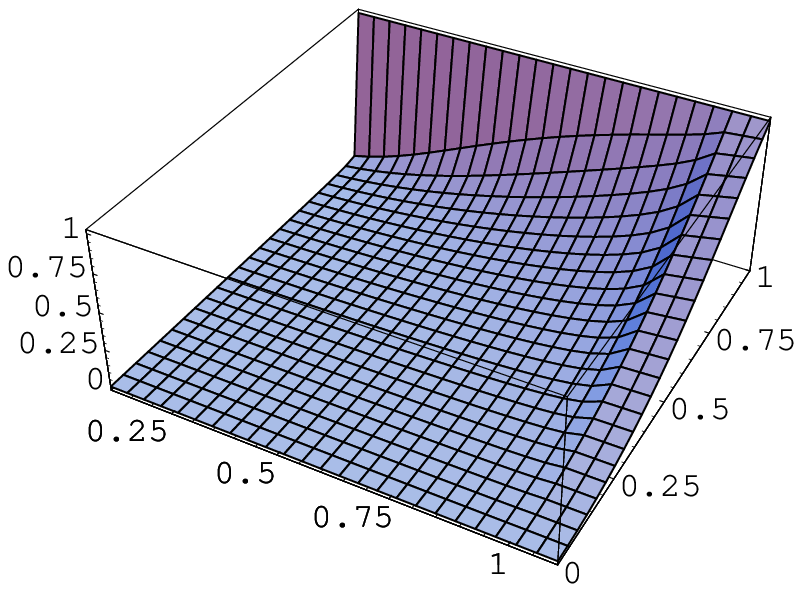}
\put(-360,-10){(a)} \put(-120,0){(b)} \put(-360,10){$\alpha$}
\put(-15,35){$p$}
\put(-460,65){$\mathcal{F}$} \put(-220,65){$\mathcal{F}$}\\
\includegraphics[width=18pc,height=12pc]{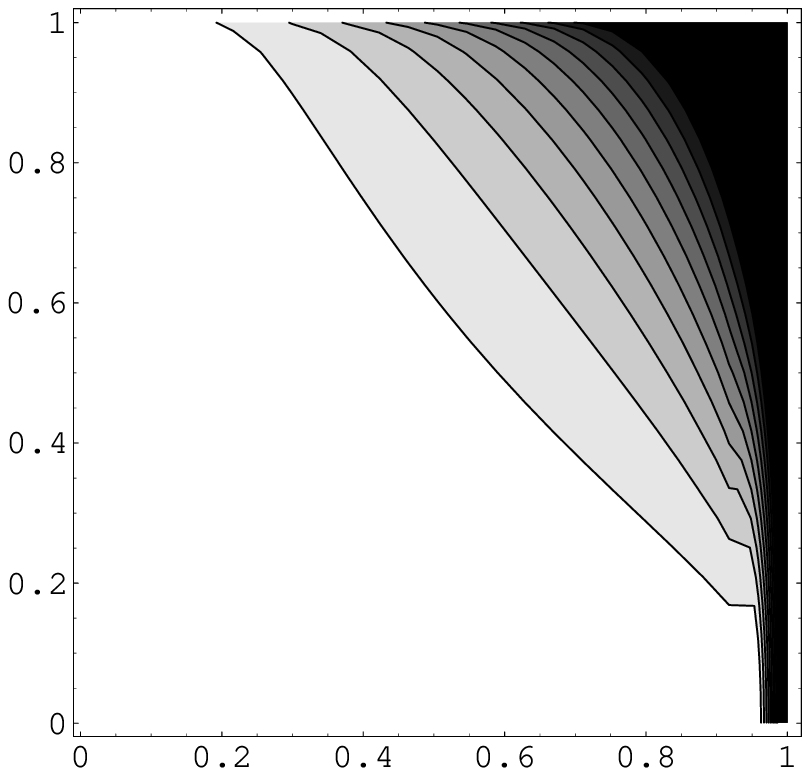}~\quad
\includegraphics[width=18pc,height=12pc]{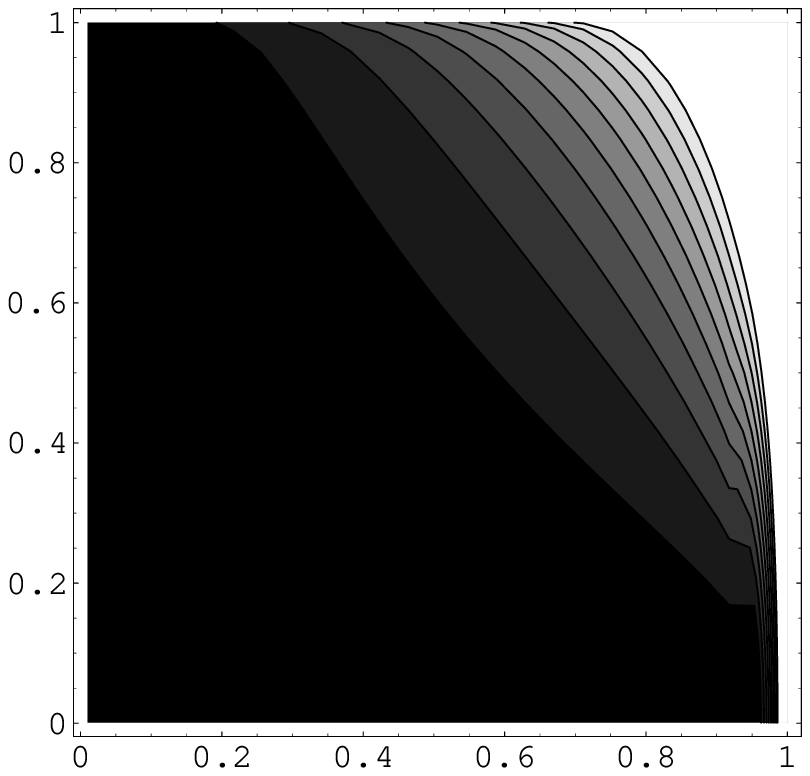}
\put(-340,-30){(c)} \put(-120,-30){(d)} \put(-330,-15){$\alpha$}
\put(10,80){$p$} \put(-460,80){$p$}
\put(-100,-15){$\alpha$}\\
\end{center}
\caption{Figs.$(a\&c)$, represent  a case of old where the
probability $|A_1|^2$ of non-infected  while  the  probability
$|B_1|^2$ for old infected is shown in  Figs.$(b\&d)$. In this
case the reserver measures $\ket{yy}\bra{yy}$ }
\end{figure}
Assume that we have a class of human cases of pathological,
described by
the neuron $\bigl|\psi _{n}\bigr\rangle=\alpha \bigl|0\bigr\rangle+\beta {%
\bigl|1\bigr\rangle}$. According to the definitions given in Table
$(2)$, this state describes a young infected with probability
$|\alpha |^{2}$ and it is non-infected with probability $|\beta
|^{2}$. In view of the friction between the two entangled tranches
of the community, the information which coded in the neuron
$\bigl|\psi _{n}\bigr\rangle$, where the virus which can be
transmitted through it, is transformed to the other segment of the
community. Also, the virus can be transmitted to another person in
the same category. Since there is an  environmental pollutions,
the transformed information interacts with these unhealthy
environment and there may be a
transfer of swine flu virus. For this example, we find with probability $25\%$%
, the virus can be transformed as shown in table(3). The values of $%
A_{i},B_{i}$ and $i=1,2,3,4$ are given by

\begin{eqnarray}
A_{1} &=&\frac{p\sqrt{(1-\alpha ^{2})}}{\sqrt{(1-q)^{2}\alpha
^{2}+p^{2}(1-\alpha ^{2})}},\quad B_{1}=\frac{(1-q)\alpha }{\sqrt{%
(1-q)^{2}\alpha ^{2}+p^{2}(1-\alpha ^{2})}}  \nonumber \\
A_{2} &=&\frac{p\alpha \sqrt{(1-\alpha
^{2})}}{\sqrt{\{1+q)^{2}(1-\alpha ^{2})^2+p^{2}\alpha
^{2}(1-\alpha ^{2})}},
\nonumber\\
\quad B_{2}&=& \frac{(1+q)(1-\alpha ^{2})}
{\sqrt{(1+q)^2(1-\alpha^2)^2+p^2\alpha^2(1-\alpha^2)}},
  \nonumber \\
A_{3} &=&\frac{(1+q)\alpha }{\sqrt{(1+q)^{2}\alpha ^{2}+p^{2}(1-\alpha ^{2})}%
},\quad B_{3}=\frac{p\sqrt{(1-\alpha ^{2})}}{\sqrt{(1+q)^{2}\alpha
^{2}+p^{2}(1-\alpha ^{2})}},  \nonumber \\
A_{4} &=&\frac{p\sqrt{(1-\alpha ^{2})}}{\sqrt{(1+q)^{2}\alpha
^{2}+p^{2}(1-\alpha ^{2})}}\quad B4=\frac{(1+q)\alpha }{\sqrt{%
(1+q)^{2}\alpha ^{2}+p^{2}(1-\alpha ^{2})}}.
\end{eqnarray}%

\begin{figure}
\begin{center}
\includegraphics[width=18pc,height=12pc]{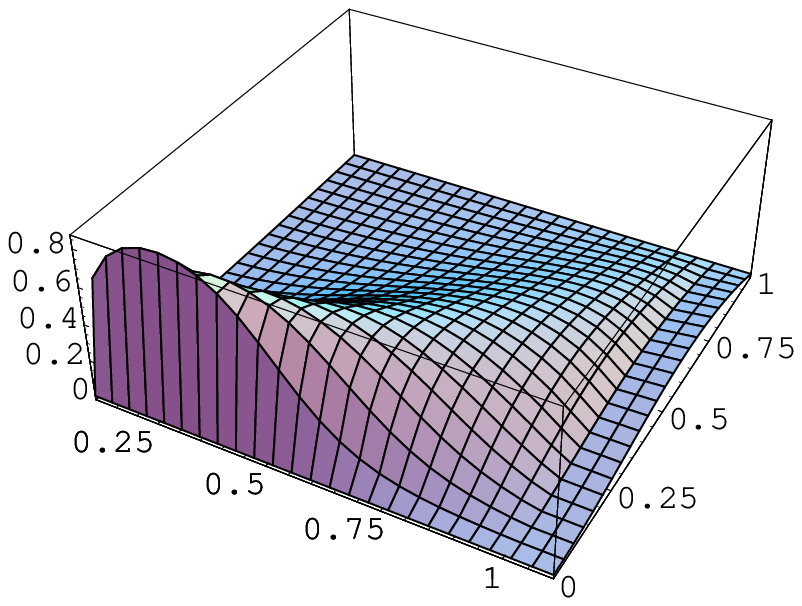}~\quad
\includegraphics[width=18pc,height=12pc]{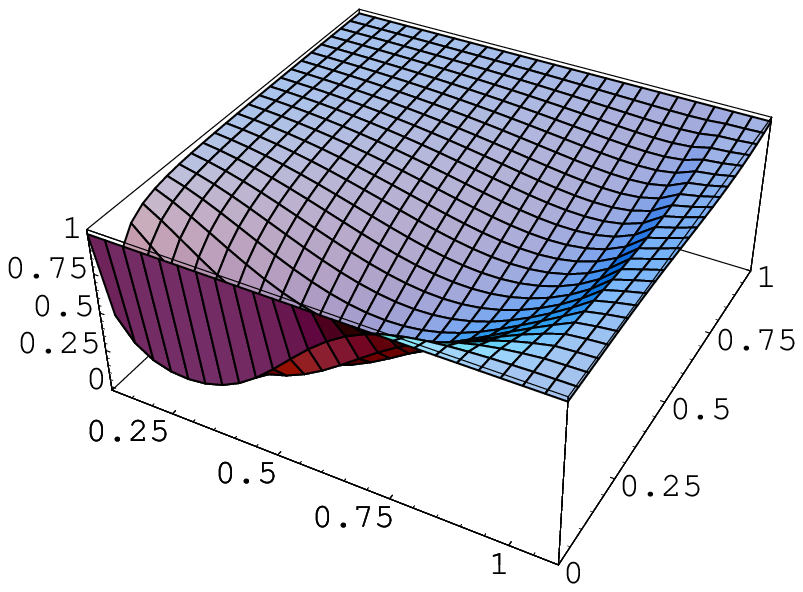}
\put(-360,-10){(a)} \put(-120,0){(b)} \put(-360,10){$\alpha$}
\put(-15,35){$p$}
\put(-460,65){$\mathcal{F}$} \put(-220,65){$\mathcal{F}$}\\
\includegraphics[width=18pc,height=12pc]{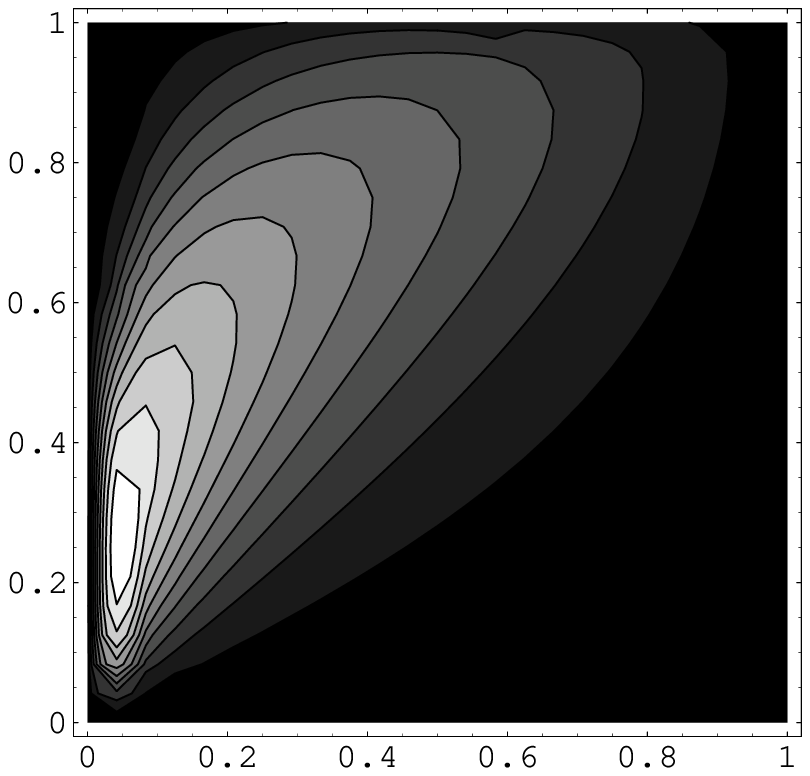} ~\quad
\includegraphics[width=18pc,height=12pc]{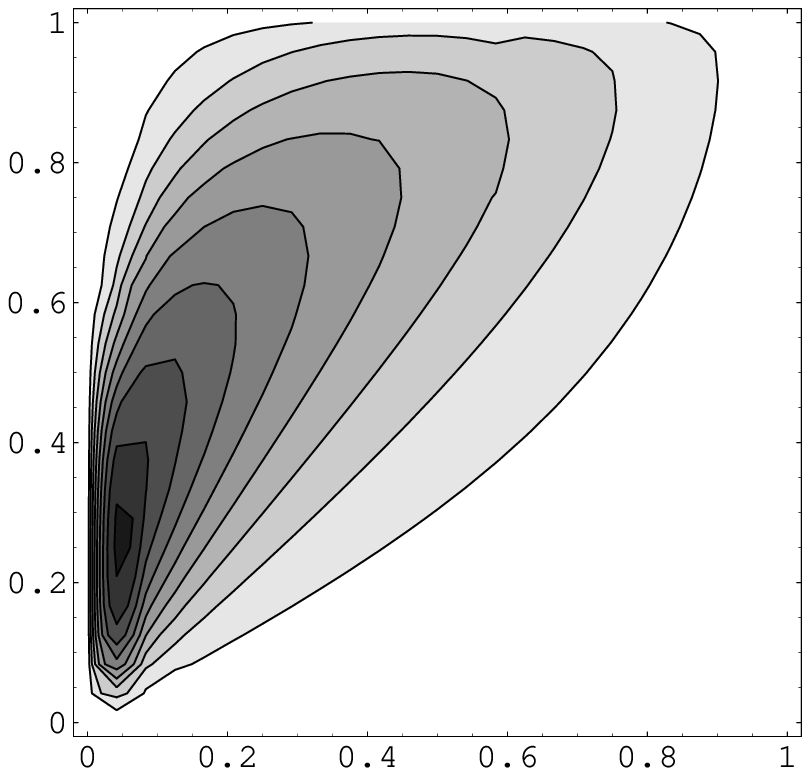}
\put(-340,-30){(c)} \put(-120,-30){(d)} \put(-330,-15){$\alpha$}
\put(10,80){$p$} \put(-460,80){$p$}
\put(-100,-15){$\alpha$}\\
\end{center}
\caption{The same as Fig.$(2)$, but for young
non-infected(Figs.$(a\&c)$) and infected (Figs.$(b\&d)$) }
\end{figure}

Table$(3)$, describes the output as  a  result of two types of
measurements that can be done by the person who receives the
transmitted information. In the
upper part of Table$(3)$, we assume that the recipient has measured $\bigl| yy %
\bigr\rangle\bigl\langle yy \bigr|$. As a result of these
measurements there are four possibilities: the first is infected
old, the second is non-infected old, the third is infected young
 and the fourth is non-infected  young. The bottom  part of
 Table$(3)$,
represents the output of the second type of
measurements, where in this case the receiver measures $\bigl| yn \bigr\rangle%
\bigl\langle yn \bigr|$. Also , one gets a similar possibilities with
different probabilities.

It is clear that, the probabilities $|A_i|^2, |B_i|^2$ for
$i=1,...4$ depend on the strength of the channel $p$, where
$p=\sqrt{1-q^2}$ and the initial state setting which is described
by the parameters $\alpha$ and $\beta$. On the other hand, the
fidelity of transmitted information  increases as the channel
parameter $p$ increases, this is clear from Fig.$(1)$. So if we
were able to control the laboratory devices to generate entangled
channel to minimize the fidelity, one can reduce the rate of
infections. Also, from theoretical point of view, if the receiver
makes some kind of rotation, one can reduce the rate of infection
between young.

In Fig.$(2)$, we plot the probabilites $A_1, B_1$ which represent
the non-infected and inflected probabilities respectively.
Fig.(2a), shows the behavior of the probability of an old
non-infected person, $|A_1|^2$. From this figure it is clear that
the person is completely non-infected i.e $|A_1|^2=1$ for large
values of the network strength. But as $p$ increases the
probability $|A_1|^2$ decreases and it becomes zero for $p\geq
0.75$. This means that for larger values of $p$ the symptoms of
the disease can  appear and its rate increases for $p\geq0.75$.
This is clear from Fig.(1b),  for $p<0.75$ and the amplitude,
$|B_1|^2=0$. From these figures we can see that the strength of
the channel  and the rate of infection playing  a central role in
the behavior of the spreading the virous.

These results are shown clearly in Fig.(1c) and Fig.(1d), where we
plot the contour diagram for the probabilities $|A_1|^2$ and
$|B_1|^2$. This description gives a clear vision for the effect of
the rate of infection  and the network strength. From these
figures, we can notice that there is gradual change from
completely dark regions to shine ones. This means that the
probability increases gradually from almost zero at (completely
dark region) to a maximum value for the completely shine regions.
As an example, in  Fig.(1c) the brightness decreases as one
increases the strength of the network, but the effect  of the
infection rate appears much earlier. This means that for large
values of $p$, the degree of entanglement of the neural network
increases and then the possibility of the infection increases. So
at $p=1$, the  probability $|A_1|^2$ is almost zero, therefore
there is an infection. Conversely, in Fig.(1D), the shine regions
increases as one  increases the network strength, $p$ and the
infection rate $\alpha$. This confirm what has been displayed in
Fig.(1b), where the probability of infection $|B_1|^2=1$  at
$p=1$.
\begin{figure}[!t]
\begin{center}
\includegraphics[width=18pc,height=12pc]{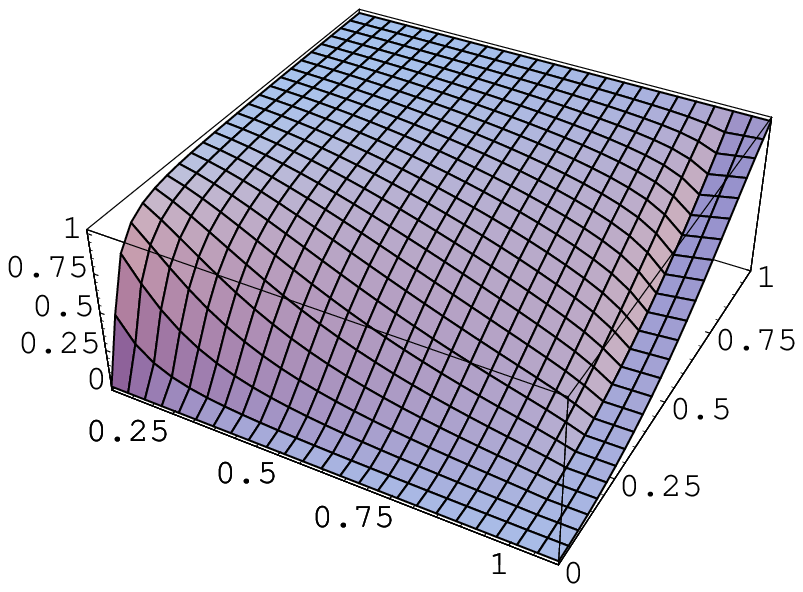}~\quad %
\includegraphics[width=18pc,height=12pc]{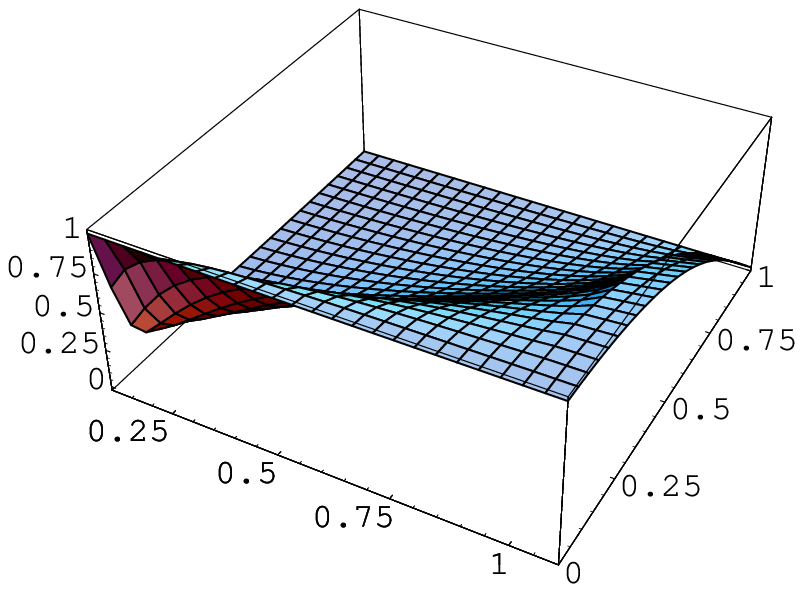}
\put(-360,-10){(a)} \put(-120,0){(b)} \put(-360,10){$\alpha$}
\put(-15,35){$p$}
\put(-460,65){$\mathcal{F}$} \put(-220,65){$\mathcal{F}$}\\
\includegraphics[width=18pc,height=12pc]{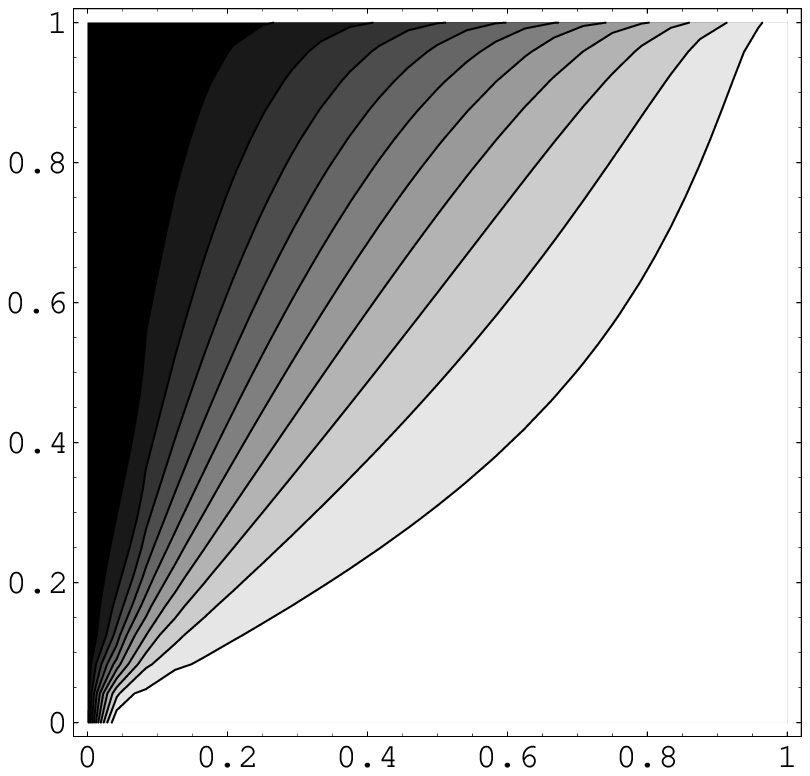} ~\quad
\includegraphics[width=18pc,height=12pc]{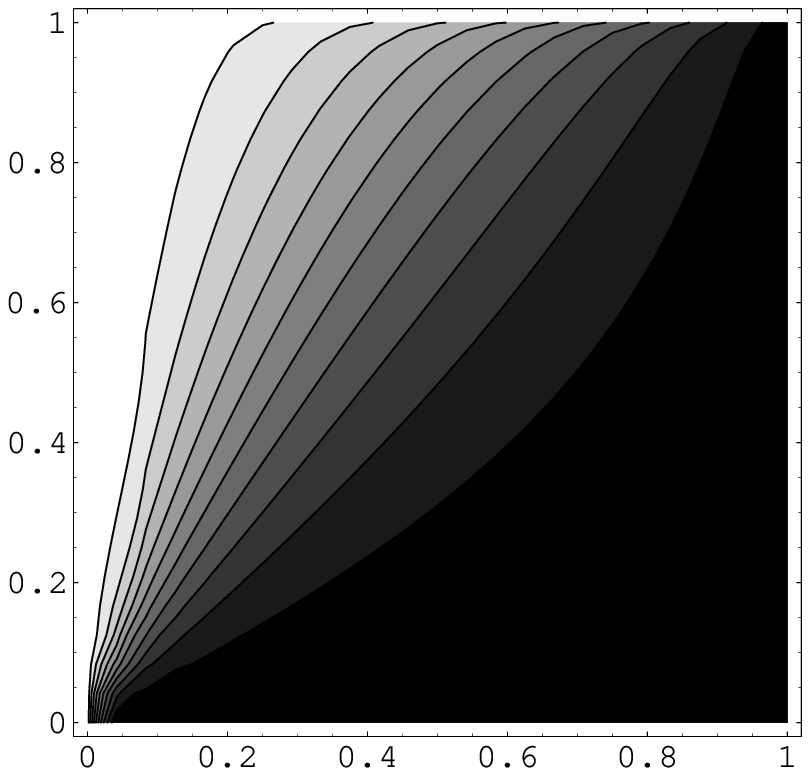}
\put(-340,-30){(c)} \put(-120,-30){(d)} \put(-330,-15){$\alpha$}
\put(10,80){$p$} \put(-460,80){$p$}
\put(-100,-15){$\alpha$}\\
\end{center}
\caption{The same as in Fig.$(2)$, but for non-infected young
(Figs.$(a\&c)$) and the infected in (Figs.$(b\&d)$) and the
receiver measures $\ket{nn}\bra{nn}$. }
\end{figure}

The other two possibilities are plotted in Fig.$3$, where the
dynamics of this amplitude represents a case of non-infected young   with probability $%
|A_2|^2$ while he is infected  with probability $|B_2|^2$. One can
see that for small values  of $\alpha$ and $p$, the young is not
infected, but the symptoms  of disease appears as one increases
the strength $p$ only. As a remark, for large values of $\alpha$
and small values of $p$, the  degree of infection is maximum (see
Fig.(3a)). Also for the probability of  infection is plotted in
Fig.(3a), where both parameters play an  important role on the
behavior of $|B_2|^2$. This behavior can be  seen clearly  from
the contour graph for each of $|A_2|^2$ and  $|B_2|^2$.

The probabilities $|A_2|^2$ and $|B_2|^2$ are displayed as contour
diagram in Fig.(3c) and Fig.(3D) respectively. In Fig.(3C), the
shining region increases for both parameters, but the dark region
is larger than the bright one. This means that the probability of
non-infected is smaller than the probability  of infected. The
converse behavior is  seen in Fig,(3d), where the dark region
(non-infected)  is very small compared with the shine regions
which represents (infection).

Finally, the behavior of the probabilities $|A_3|^2$ and $|B_3|^2$
are shown in Fig.$(4)$, where the receiver measures
$\ket{yn}\bra{yn}$. In Fig.(4a), we plot the probability  of a
non-infected young, $|A_3|^2$. This figure shows that the
probability increases  the network strength increases  and it is
almost zero for small values. This means that for small values of
$p$, the probability that the young is infected is increases. This
phenomena is shown in Fig,.$(4b)$, where $|B_2|^2=1$ as one
decreases the network's strength. The confirmation of these
results is shown in Figs.$(4c\&4d)$, where the dark regions appear
for a small range of $\alpha$  and $p$ as shown in Fig.$(4c)$,
while it  increases for a large range of these parameters. Finally
concerning the last possibility which for old non-infected is
represented  by the probability $|A_4|^2$ and infected old
 with probability $|B_4|^2$, we can see that they have the
same behavior of $|B_3|^2$ and $|A_3|^2$. In other words, the
probability for non-infected old  equals to the probability of
infected young and the probability that old infected equals the
non-infected young. So, we obtain the same behavior and there is
no need to re-plot them.

\section{Conclusion}

In this contribution, we employ the quantum neural network to achieve the
quantum teleportation protocol. In our treatments we assume that the neurons
are connected with unperfect channels. The sensitivity of the fidelity of
the transmitted information is investigated for the network's parameter as well as for
 the structure of the input information. For
large values of the channel parameter, the channel becomes a
maximum entangled and consequently the fidelity is maximum. The
idea of quantum teleporation is applied on  a practical example,
where we investigate the possibility of controlling on the spread
of swine flu virus. In this context, we show that the virous can
be transformed to a similar  segment of society with different
probabilities. Also, it has been shown, that the case of patients
has no effect on the spread of the virus for some possibility. For
other possibility the degree of infection plays the central role,
while for some other cases both parameters paly an equivalent
role. On the other hand, by local rotations young people can avoid
HIN1 infection.

We note that in this protocol we used a quantum neural network consists of
two neurons. We believe that there are many factors affecting the spread of
swine flu virus. Therefore, we can develop this protocol to include many of
the human factors and external factors, and then strategies can be developed
to understand the process of transmission. In fact, this protocol was an
attempt to understand the theory of the spread of the virus as an
application of quantum neural networks. And we hope that this study
contributes to a vision to reduce the spread of the epidemic.

\bigskip \textbf{Acknowledgements:} The designated project has been
fulfilled by financial support from the unite research of
IT-House, Aswan, Egypt.

\end{document}